\documentclass[]{aa} % for the letters 

\usepackage{graphicx}
\usepackage{txfonts}
\begin{document}

   \title{The Musca cloud: A 6~pc-long velocity-coherent, sonic filament \thanks{This publication is based on data acquired with the Atacama Pathfinder Experiment (APEX). APEX is a collaboration between the Max-Planck-Institut fuer Radioastronomie, the European Southern Observatory, and the Onsala Space Observatory (ESO programme 087.C-0583).}
}
  
   %\subtitle{I. Overviewing the $\kappa$-mechanism}

   \author{A. Hacar
          \inst{1}
          \and
          J. Kainulainen\inst{2}
          \and
          M. Tafalla\inst{3}
          \and
          H. Beuther\inst{2}
          \and
          J. Alves\inst{1}
          }

   \institute{Department of Astrophysics, University of Vienna,
              T\"urkenschanzstrasse 17, A-1180 Vienna, Austria\\
              \email{alvaro.hacar@univie.ac.at}
         \and
             Max-Planck Institute f\"ur Astronomie, K\"onigstuhl 17, 69117 Heidelberg, Germany
         \and
             Observatorio Astronomico Nacional (OAN-IGN), Alfonso XII 3, 28014, Madrid, Spain
             }

   \date{XXX-XXX}

% \abstract{}{}{}{}{} 
% 5 {} token are mandatory

  \abstract{Filaments play a central role in the molecular clouds' evolution, but their internal dynamical properties remain poorly characterized. To further explore the physical state of these structures, we have investigated the kinematic properties of the Musca cloud. We have sampled the main axis of this filamentary cloud in $^{13}$CO and C$^{18}$O (2--1) lines using APEX observations. The different line profiles in Musca shows that this cloud presents a continuous and quiescent velocity field along its $\sim$~6.5~pc of length. With an internal gas kinematics dominated by thermal motions (i.e. $\sigma_{NT}/c_s\lesssim1$) and large-scale velocity gradients, these results reveal Musca as the longest velocity-coherent, sonic-like object identified so far in the ISM. 
  The transonic properties of Musca present a clear departure from the predicted supersonic velocity dispersions expected in the Larson's velocity dispersion-size relationship, and constitute the first observational evidence of a filament fully decoupled from the turbulent regime over multi-parsec scales.
  }

   \maketitle
%
%________________________________________________________________

\section{Introduction}
 
Since the earliest molecular line observations, it is well established that the internal gas kinematics of molecular clouds is dominated by supersonic motions \citep[e.g.,][]{ZUC74}. These motions give rise to an empirical correlation between the velocity dispersion and the size-scale in molecular clouds, the so-called Larson's relation \citep{LAR81}. In our prevalent paradigm of the turbulent ISM \citep[e.g.,][]{MAC07}, this correlation arises from the power-law scaling of kinetic energy expected in a Kolmogorov-type cascade for a turbulence dominated fluid. As part of the turbulence decay, the density fluctuations created as a result of the supersonic gas collisions at distinct scales are responsible for the formation of the internal substructure of these objects \citep[see ][for a review]{ELM04}. 
 
Identifying the first (sub-)sonic structures formed inside molecular clouds is of fundamental importance to understanding their internal evolution. In the absence of supersonic compressible motions, the maximum extent of these sonic regions defines the end of the turbulent regime and the transitional scales at which turbulence ceases to dominate the structure of the cloud.
From the analysis of the gas velocity dispersion in molecular line observations, \citet{GOO98} and \citet{PIN10} identified the dense cores as the first sonic structures decoupled from the turbulent flow, typically on scales of $\sim$~0.1~pc. More recently, \citet{HAC11} have shown that these dense cores are embedded in distinct $\sim$~0.5~pc length, velocity-coherent filamentary structures, referred to as fibers, with identical sonic-like properties to those previously measured in cores. According to these latest results, the presence of this sonic regime precedes the formation of cores extending up to (at least) subparsec scales. 

The interpretation of the supersonic motions reported in more massive filaments at larger scales is, however, matter of a vigorous debate.
The ubiquitous presence of filaments in both star-forming and pristine molecular clouds revealed by the latest Herschel results indicates that their formation is particularly favored as part of the turbulent cascade \citep{AND10,AND14}. From the comparison of different clouds, \citet{ARZ13} has suggested that the internal velocity dispersion of those gravitationally bound filaments increases with their linear mass as a result of their gravitational collapse. On the other hand, \citet{HAC13} have demonstrated that supercritical filaments like the 10~pc long \object{B213}-\object{L1495} region in Taurus are actually complex bundles of fibers. In these bundles, the apparent broad and supersonic linewidths are produced by the line-of-sight superposition of multiple and individual sonic-like structures at distinct velocities. New observations are needed to clarify this controversy.

\begin{figure}
   \centering
   \includegraphics[width=\columnwidth]{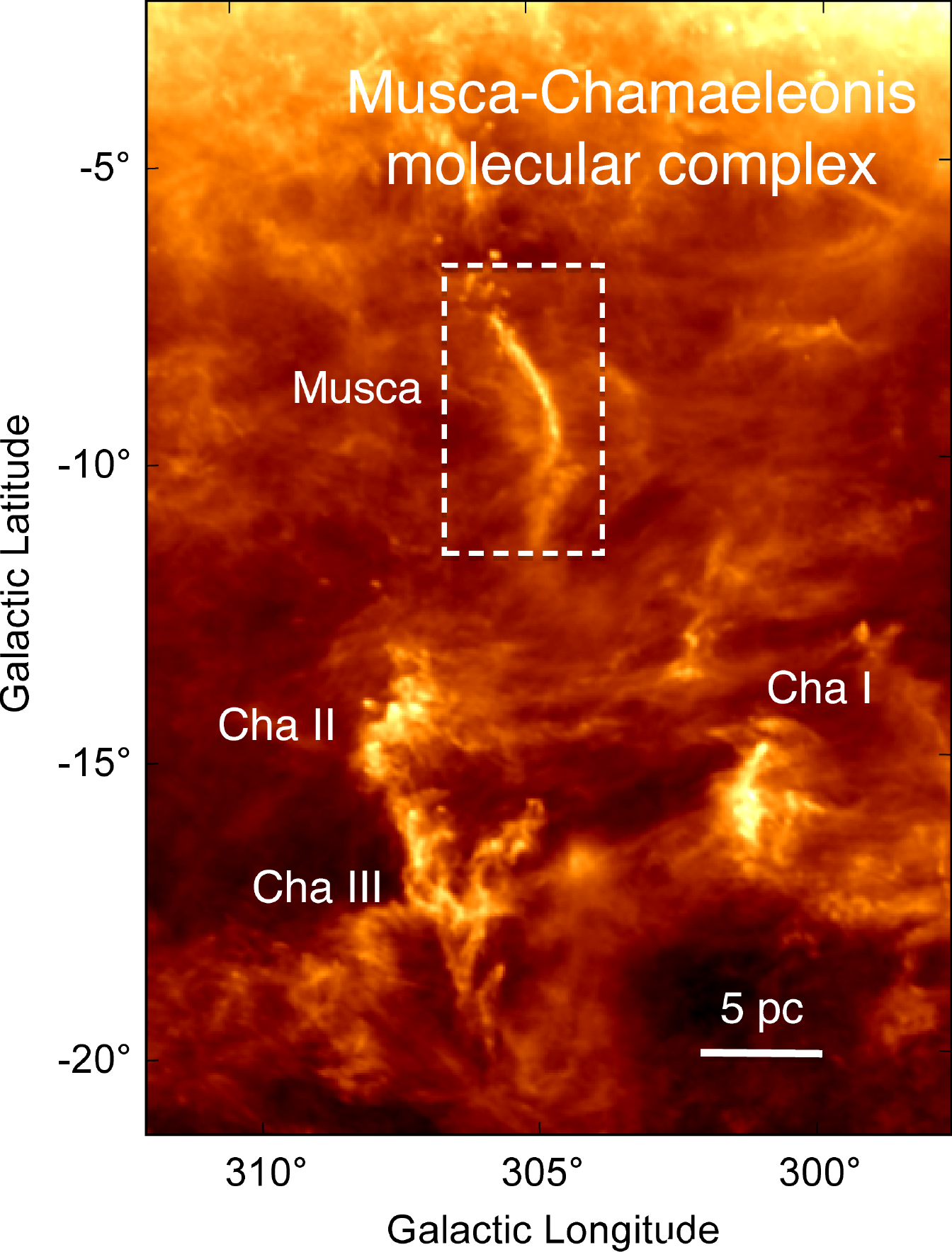}
   \caption{Planck 857~GHz emission map, oriented  in Galactic coordinates, of the northern part of the Musca-Chamaeleonis molecular complex. The different subregions are labeled in the plot. Note the favourable properties of the Musca filament, showing a quasi-rectilinear geometry along its $\sim$6~pc of length, perfectly isolated from the rest of the complex.}
              \label{fig:map}%
    \end{figure}

In this paper we report the analysis of the internal gas kinematics of the \object{Musca} cloud $(\alpha$,$\delta)_{J2000}=(12^h23^m00\fs0,-71\degr20\arcmin00\arcsec$) as a paradigmatic example of a pristine and isolated filament.
With an estimated distance of $\sim$~150~pc derived from extinction measurements \citep{FRA91}, Musca is located at the northern end of the Musca-Chamaleonis molecular complex. As a characteristic feature, Musca exhibits an almost perfect rectilinear structure along its total length of more than 6~pc (see  Fig.~\ref{fig:map}). 
\citet{KAI09} showed that the column density distribution of Musca has a shape more typical for star-forming than non-star-forming clouds. Yet, the cloud contains only a few solar masses at densities capable for star formation \citet{KAI14b}, suggesting that it may be in transition between quiescence and star formation \citep[cf.][]{KAI14}.
Indeed, early studies of this cloud have reported the presence of a single embedded object, \object{IRAS12322-7023} (a TTauri candidate source), and a handful of dense but still prestellar, cores \citep{VIL94,JUV12}.
Optical \citep{ARN93,PER04} and dust polarization \citep{PLA14} measurements have revealed the striking configuration of the magnetic field in this region which is perpendicularly oriented to the main axis of this filament. Large-scale extinction and submillimeter continuum maps indicate that its mass per unit length is similar to the value expected for a filament in hydrostatic equilibrium \citep{KAI14}. Additionally, millimeter line observations along this cloud have reported some of the narrowest lines detected in molecular clouds \citep{ARN93,VIL94}. 
These extraordinary physical properties make Musca an ideal candidate to explore the dynamical state of a filament in its early stages of evolution.

%__________________________________________________________________

\section{Molecular line observations}

%______________________________________________

Between May and June 2011, we surveyed a total of 300 independent positions along the Musca cloud using the APEX12m telescope.
As illustrated in Fig.~\ref{fig:spectra} (left), the observed points correspond with a longitudinal cut along the main axis of the Musca filament. 
 Designed to optimize the observing time on source, this axis follows the highest column density crest of this filament and
the most prominent features identified in both extinction and continuum maps \citep{KAI14}.
This axis was sampled every $\sim$~30~arcsec, corresponding to a typical separation of approximately one beam at the frequency of 219 GHz ($\Theta_{mb}=28.5$ arcsec). 

In order to cover both transitions simultaneously, the APEX-SHeFI receiver was tuned to the intermediate frequency between the $^{13}$CO (J=2--1) \citep[$\nu=$~220398.684 MHz; ][]{CAZ04} and C$^{18}$O (J=2--1) \citep[$\nu=$~219560.358 MHz; ][]{CAZ03} lines. All the observations were carried out in Position-Switching mode using a similar OFF position for all the spectra with coordinates $(\alpha$,$\delta)_{J2000}=(12^h41^m38\fs0,-71\degr11\arcmin00\arcsec$), selected from the large-scale extinction maps of \citet{KAI09} as a position presenting extinctions values of A$_V<$~0.5$^{mag}$.
To improve the observing efficiency, and in collaboration with the APEX team, a new observing technique was developed for this project where each group of 3 individual positions (not necessarily aligned) shared a single OFF integration, similar to the method using in raster maps. The typical integration time per point was set to 2 min on-source, while all the observations were carried out under standard weather conditions (PWV~$<$~3mm). Pointing and focus corrections and line calibrations were regularly checked every 1-1.5 hours.

As results of an instrumental upgrade in June 2011, two distinct backends were used for this project. 
Approximately half of the positions were surveyed using the now decommissioned Fast Fourier Transform Spectrometer (FFTS) with an effective spectral resolution of 122~KHz or $\sim$~0.17 km~s$^{-1}$ at the central frequency of 220~GHz. The second half of these observations were carried out using the new facility RPG eXtended bandwidth Fast Fourier Transform Spectrometer (XFFTS) backend with an improved resolution of 76~kHz or $\sim$~0.10 km~s$^{-1}$ at the same frequencies. To check the consistency between the two datasets, eight of the most prominent positions along the filament were observed in both configurations. A systematic difference of 80~kHz (or $\sim$~0.10 km~s$^{-1}$) was found between the signal detected in both FFTS and XFFTS backends, which was attributed to hardware problems in the previous FFTS installation (C. DeBreuck, private communication). This frequency correction was then applied to the final sample of FFTS data.

The final data reduction included a combination of the two datasets, where the XFFTS spectra were smoothed and resampled into the FFTS resolution in those positions were both observations were available, and a third-order polynomial baseline correction using the CLASS software\footnote{http://www.iram.fr/IRAMFR/GILDAS}. To achieve this, each individual $^{13}$CO and C$^{18}$O spectrum was reduced independently. According to the facility-provided antenna parameters included in the APEX website\footnote{http://www.apex-telescope.org/telescope/efficiency/index.php}, a final intensity calibration into main-beam temperatures was achieved by applying a beam efficiency correction of $\eta_{mb}=$~0.75. The resulting spectra present a typical rms value of 0.14~K.

\section{Data analysis}\label{sec:fits}

\subsection{Data overview}

%______________________________________________
   \begin{figure*}
   \centering
   \includegraphics[width=\textwidth]{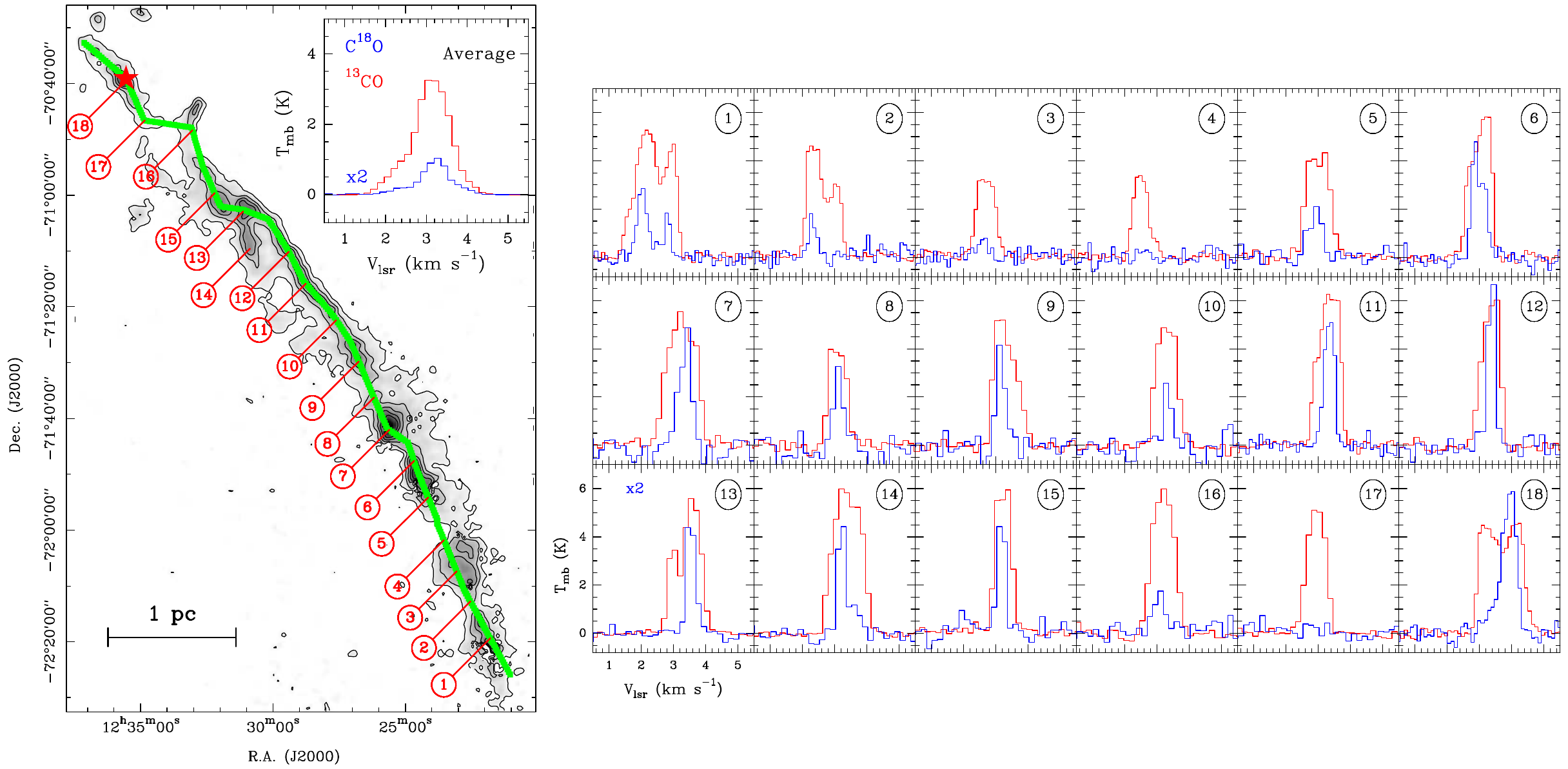}
   \caption{(Left) Total column density map derived from NIR extinction measurements \citep{KAI14}. The contours are equally spaced every 2 mag in A$_V$. The 300 positions surveyed by our APEX observations in both the $^{13}$CO and C$^{18}$O (2--1) lines are marked in green following the main axis of this filament. The red star in the upper left corner denotes the position of the TTauri IRAS12322-7023 source. (Right) $^{13}$CO (2-1) (red) and C$^{18}$O (2-1) (blue; multiplied by 2) line profiles found in different representative positions along this cloud (spectra 1-18), all labeled on the map. The averaged spectra of all the data available inside this region are also shown in the upper right corner of the extinction map.}
              \label{fig:spectra}%
    \end{figure*}
%

% Peaks, data description
Figure~\ref{fig:spectra} (right) illustrates several representative $^{13}$CO and C$^{18}$O (2-1) spectra found along Musca.
As seen in the corresponding average spectra (upper right corner, in the left panel), the total emission of both lines appears at velocities between $\sim$~1.5 and 4.5 km~s$^{-1}$ with a smooth and continuous variation along the total $\sim$~6~pc of this filament (see spectra 1-18). Most of the observed positions present narrow lines with a prevalence of single-peaked spectra in both CO isotopologues. The observed velocity variations are consistently traced in both CO lines (see also Sect.\ref{sec:results}). These kinematic properties reproduce the results obtained in the $^{13}$CO and C$^{18}$O (1-0) observations carried out by \citet{VIL94} at different positions along Musca. Compared to the limited sample of 16 points studied by these authors, however, our  survey of 300 observations allows us to investigate the dynamical state of this filament in detail.

A simple inspection of the individual line profiles found along the Musca filament reveals some of its particular kinematic features. 
Some C$^{18}$O (2-1) spectra present total linewidths of less than two channels  (e.g. spectra 12 and 15 in Fig.~\ref{fig:spectra}), being unresolved by our high spectral resolution observations.  Additionally, most of these observed positions present a single component in velocity. Three well-defined regions contain most of the double-peaked spectra, typically detected in $^{13}$CO. The most prominent one is located at the northern end of Musca, coincident with the position of the IRAS12322-7023 source. The emission appears as a double-peaked line in $^{13}$CO and typically as a single line with a blue-shifted wing in C$^{18}$O (see spectrum 18). Restricted to the vicinity of this IRAS source, this complex kinematics appears to be related to the interaction of this embedded object with its envelope.

 Most of the double-peaked spectra detected in C$^{18}$O are preferentially located at the southern end of Musca extending throughout a region of about 10~arcmin in length. In this case, two velocity components are detected in both $^{13}$CO and C$^{18}$O isotopologues with roughly similar line ratios (see spectrum 1) suggesting a superposition of a secondary component along this region. A much clearer example of this behavior is found in the third of these regions, located at the intersection of what appears to be two independent branches identified in the central part of the Musca filament according to our extinction maps \citep[][ see also spectra 13 and 14]{KAI14}. 

This simple kinematic structure contrasts with the high level of complexity found in more massive bundle-like structures like B213-L1495 that ahve spectra characterized by high-multiplicity and high spatial variability \citep{HAC13}. Opposite to it, Musca resembles some of the quiescent properties of the single velocity-coherent fibers observed at scales of $\sim$~0.5~pc in regions like L1517 \citep{HAC11}. Compared to these last objects, the quiescent nature of Musca seems to extend up to scales comparable to the total length of this cloud. In addition to its extraordinary rectilinear geometry, these observations suggest that Musca is the simplest molecular filament ever observed.

\subsection{Gaussian decomposition}

Following the analysis techniques used by \citet{HAC11} and \citet{HAC13}, we parametrized all the kinematic information present in our $^{13}$CO and C$^{18}$O data by fitting Gaussians to all our spectra using standard CLASS routines. For that purpose, each individual spectrum was examined and fitted independently. In most cases, one single component was used at each individual position. A maximum of two independent components were fitted only if the spectrum presented a clear double-peaked profile whose two maxima were separated by at least three channels in velocity (e.g. spectrum 1 in Fig.~\ref{fig:spectra}). In case of doubt (i.e. top-flat or wing-like spectra; e.g. spectrum 5), one line component was fitted. This conservative approach was preferred to prevent the split of the line emission into artificially narrow subcomponents.

From the total of 300 spectra fitted for each line, $\sim$~80\% of the $^{13}$CO and 95\% of the C$^{18}$O line profiles were identified and fitted as single-peak spectra with S/N~$\geq$~3. The other 20\% in the case of $^{13}$CO and 5\% of the C$^{18}$O spectra were fitted with two independent components. Most of the fitted profiles, either in single or multiple spectra, appear to be well reproduced by Gaussian components.  
With perhaps the exception of the wing-like emission around IRAS12322-7023 (i.e. spectrum 18), our Gaussian decomposition seems to capture the main kinematic properties of the gas traced by the two CO lines studied in this paper.

\section{Results}

\subsection{Gas properties: temperature, column density, and molecular abundances}\label{sec:tracers}

%______________________________________________ 
   \begin{figure*}
   \centering
   \includegraphics[width=\textwidth]{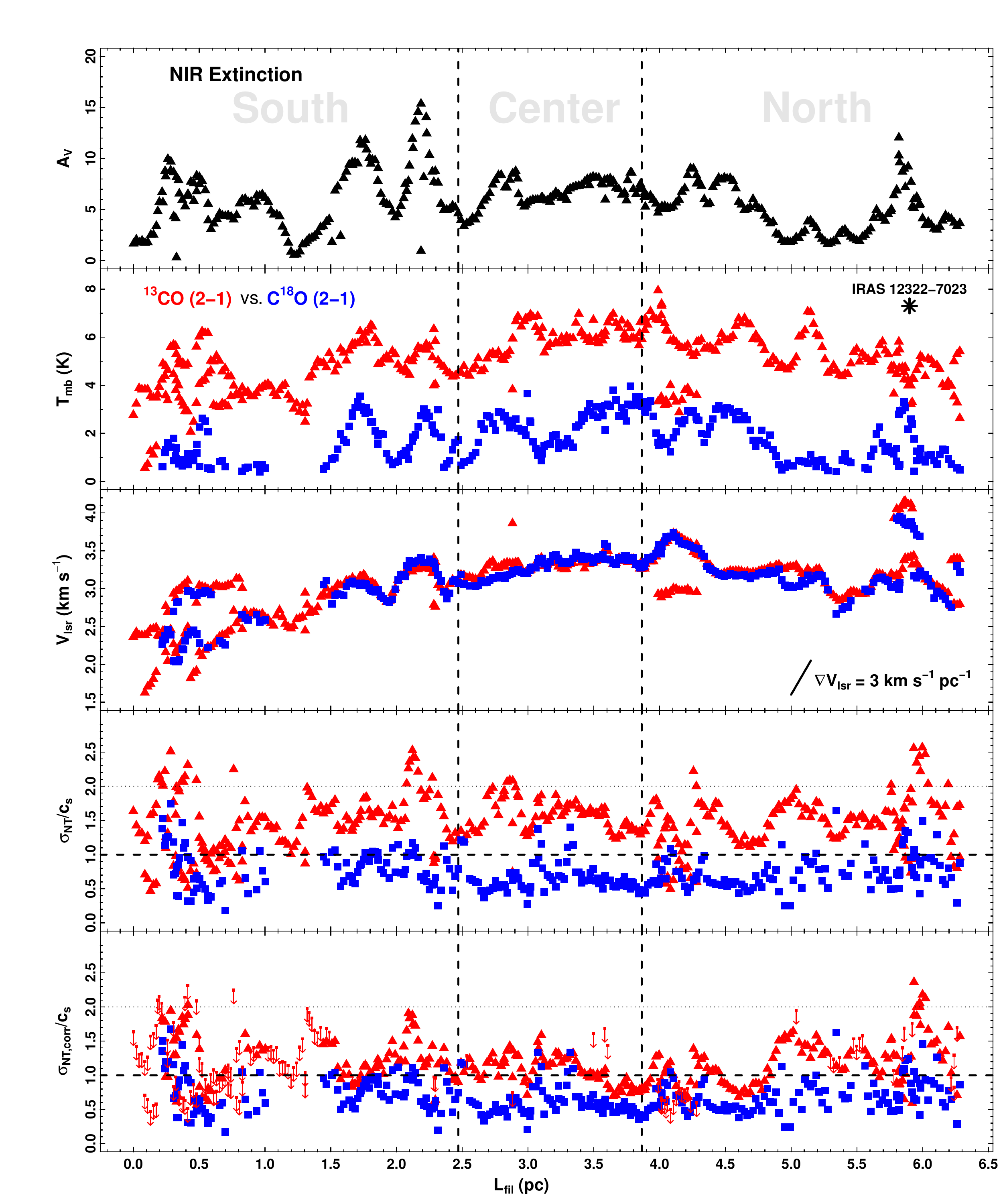}
   \caption{Observational results combining both NIR extinction (black) and both $^{13}$CO (red triangles) plus C$^{18}$O (blue squares) (2-1) line measurements along the main axis of Musca. From Top to Bottom: (1) Extinction profile obtained by \citet{KAI14}; (2) CO main-beam brightness temperatures (T$_{mb}$); (3) centroid velocity (V$_{lsr}$); (4) non-thermal velocity dispersion along the line of sight ($\sigma_{NT}$); and (5) opacity corrected non-thermal velocity dispersion along the line of sight ($\sigma_{NT,corr}$) including upper limits for the $^{13}$CO emission (red arrows; see text for discussion). The velocity dispersion measurements are expressed in units of the sound speed c$_s=0.19$~km~s$^{-1}$ at 10~K. In these last plots the horizontal lines delimitate to the sonic ($\sigma_{NT}/c_s\leq1$; thick dashed line) and transonic ($\sigma_{NT}/c_s\leq2$; thin dotted line) regimes, respectively. Only those components with a S/N~$\geq$~3 are displayed in the plot. The position of the IRAS 12322-7023 source as well as a typical 3 km~s$^{-1}$~pc$^{-1}$ velocity gradient are indicated in their corresponding panels. We note the presence of the double-peaked spectra at the positions L$_{fil}\sim$~0.4,4.2, and 5.8~pc in the middle panel. The two vertical lines delimitate the south, central, and north subregions identified by \citet{KAI14}.
   }
              \label{fig:kinematics}%
    \end{figure*}
%______________________________________________ 

Owing to the observed differences in their integrated intensity maps, the most abundant CO isotopologues are  commonly interpreted as density selective tracers: $^{12}$CO and $^{13}$CO seem to trace the most diffuse gas in clouds while only C$^{18}$O is detected in regions with highest column- and volume densities. This interpretation relies on the oversimplified assumption that the expected fractional abundances of these molecular tracers (e.g. X(C$^{18}$O):X($^{13}$CO):X($^{12}$CO) = 1:7.3:560 \citet{WIL94}) are translated into differential sensitivity thresholds for the distinct density regimes in clouds. Physically speaking, however, all the CO isotopologues share quasidentical properties in terms of their excitation conditions and chemistry. If secondary effects like radiative trapping \citep{EVA99} and photodissociation within self-shielded regions \citep[A$_V>$~2-3;][]{VDH88} are neglected, these molecular tracers should be sensitive to roughly the same gas at all densities and along the same line of sight. Observationally, this picture is complicated by the different emissivities and optical depths of their lower-J rotational transitions. Thanks to their large relative abundances, the detection of the most abundant isotopologues (i.e., $^{12}$CO and $^{13}$CO) is favored over their rarest counterparts (e.g., C$^{18}$O or C$^{17}$O) particularly in the diffuse outskirts of the clouds. In addition, these large relative abundances also produce strong variations in their line opacities (i.e., $\tau(A)\sim X(A/B)\cdot \tau(B)$ for the same J-transition; \citet{MYE83a}). As a result, tracers like $^{12}$CO or, to lesser extent, $^{13}$CO become heavily saturated in regions with high column densities.

% Temperature and opacity
The combination of multiple CO lines can be used to derive different physical properties of the molecular gas inside clouds \citep[e.g.,][]{MYE83a,VIL94}. The three main parameters derived from our Gaussian fits with S/N~$\geq$~3 (i.e., peak line temperature, velocity centroids, and velocity dispersion) are presented in Fig.~\ref{fig:kinematics} as a function of the position along the main axis of the Musca filament (L$_{fil}$) measured from the southern end of this cloud. For comparison, this figure (first panel) includes the near-IR (NIR) extinction values derived along the same axis \citep{KAI14}. The first of the parameters studied there corresponds with the observed line intensity or brightness temperature T$_{mb}$ (second panel).  Values for T$_{mb}$ of 5-7~K and 1-3~K are found for the $^{13}$CO and C$^{18}$O (2-1) components, respectively. Compared to the canonical line ratio of 7.3 expected in the optically thin case, a point-to-point comparison shows characteristic values of T$_{mb}(^{13}$CO$)/$T$_{mb}($C$^{18}$O$)\sim 5$, indicative of moderate opacities for the $^{13}$CO (2-1) lines  (see below). In the optically thick approximation, the observed line intensities lead to excitation temperatures of T$_{ex}(^{13}CO)=$~9-11~K. These results are in close agreement with the average gas kinetic temperatures found in starless cores \citep[T$_K\sim$10~K;][]{BEN89} and the dust effective temperatures reported for the two most prominent starless cores observed in Musca \citep[T$_{dust}$=11.3~K;][]{JUV10} (cores 4 and 5 according to \citet{KAI14} notation). For practical purposes, in this paper we have assumed LTE excitation conditions with a uniform gas kinetic temperature of T$_K=10$~K for all the gas traced by our CO observations (i.e. $\mathrm{T}_{ex}(^{13}\mathrm{CO})=\mathrm{T}_{ex}(\mathrm{C}^{18}\mathrm{O})=10$~K).

\begin{figure}
   \centering
   \includegraphics[width=\columnwidth]{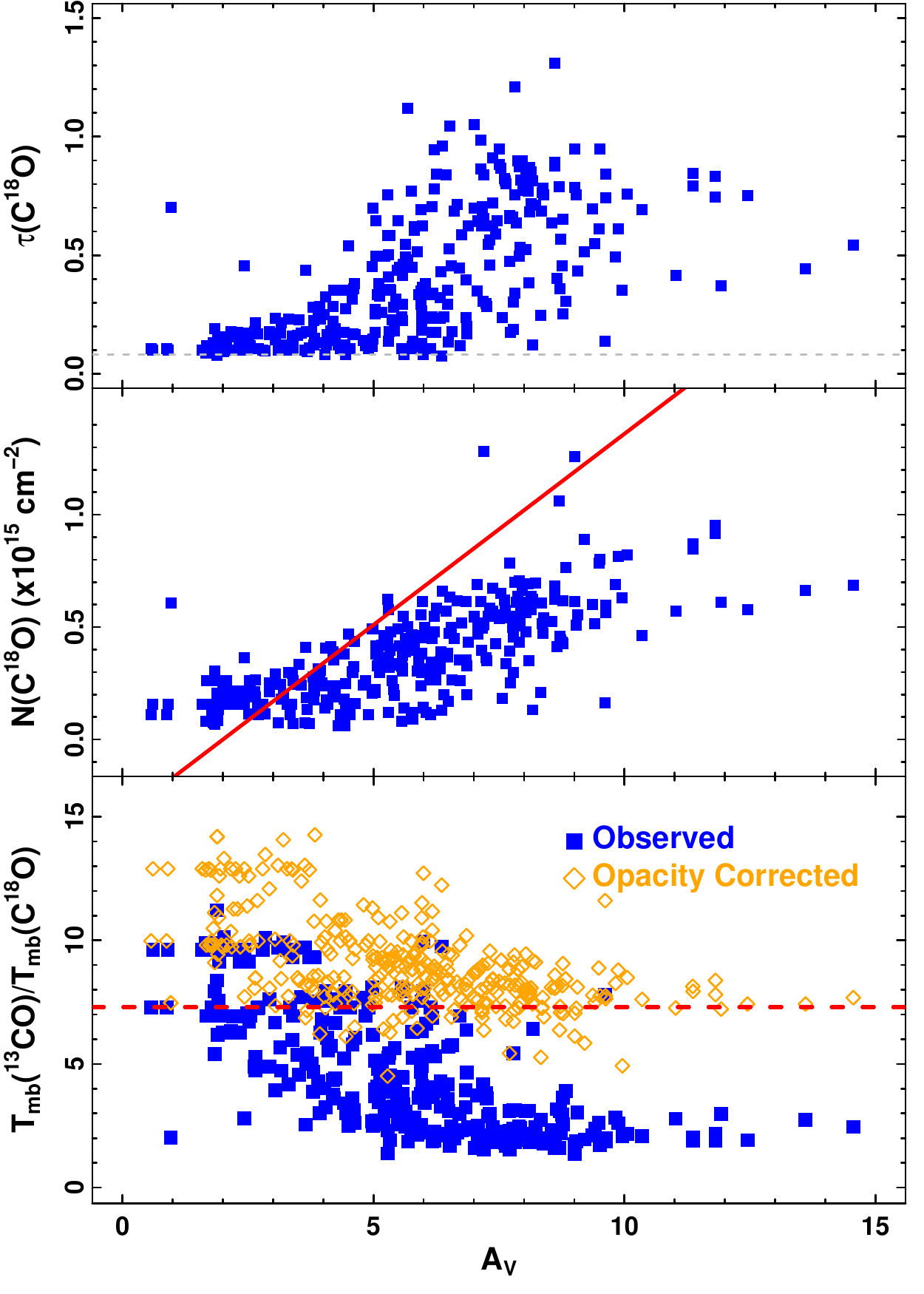}
   \caption{Observational properties of the observed C$^{18}$O and $^{13}$CO (2-1) lines as a function of the total column density observed along the Musca filament: (Upper panel) Optical depth of the C$^{18}$O (2-1) lines; (Mid panel) C$^{18}$O total column density; (Lower panel)  Observed (blue solid squares) and opacity corrected (orange diamonds) T$_{mb}(^{13}$CO$)/$T$_{mb}($C$^{18}$O$)$ line ratio. The gray dashed line in the Upper panel corresponds to the characteristic detection S/N=3 line detection threshold for a spectrum with an rms value of 0.14~K, in T$_{mb}$ units.
The derived C$^{18}$O column densities are consistent (within a factor of two) with the typical abundances obtained in clouds like Taurus and Ophiuchus  \citep[solid red line in the Mid panel;][]{FRE82}. The red horizontal line in the lower panel indicates the canonical T$_{mb}(^{13}$CO$)/$T$_{mb}($C$^{18}$O$)$=7.3 line ratio expected in an optically thin regime. }
              \label{fig:abundances}%
    \end{figure}

At each position observed along the Musca cloud, we have estimated the central optical depth $\tau_0$ of the C$^{18}$O (2-1) emission from the peak temperatures measured in our spectra as:
\begin{equation}\label{eq:tau0}
	\tau_0(C^{18}O) = -log \left( 1- \frac{T_{mb}(C^{18}O)}{J(T_{ex})-J(T_{bg})} \right)
\end{equation}
obtained solving the radiative transfer equation where $J(T)=\frac{h\nu/k}{exp(h\nu/kT)-1}$ \citep[e.g.,][]{TOOLS}.
Figure ~\ref{fig:abundances} presents the linear increase of the resulting C$^{18}$O (2-1) line opacity as a function of the extinction values reported by \citet{KAI14} at the same positions.
At those column densities where this molecule is detected, we obtain opacity values of $\tau($C$^{18}$O(2-1))=0.08-1.3. 
Assuming standard CO fractionation, this optically thin C$^{18}$O emission yields optically thick estimates of $\tau(^{13}$CO(2-1))=0.5-9.5 and, although not observed here, $\tau(^{12}$CO(2-1))$\sim30-400$.

The different line opacities of the $^{13}$CO and C$^{18}$O lines are recognized in several of the parameters derived in our fits. Figure~\ref{fig:abundances} (Mid panel) presents the evolution of the C$^{18}$O column densities, estimated in LTE conditions from the integrated intensity of these lines as a function of A$_V$.  As expected for an optically thin tracer with constant abundance, a linear correlation is recovered between these two observables.    
In absolute terms, the CO abundances derived from the C$^{18}$O column densities obtained in Musca are consistent, within a factor of 2, with the values found in clouds like Taurus and Ophiuchus \citep[i.e., X(C$^{18}$O/H$_2$)$\sim 1.7\times 10^{-7}$; see Fig.\ref{fig:abundances};][]{FRE82}. 
Instead, the increasing line opacity of the $^{13}$CO lines is reflected in the monotonic decrease of the T$_{mb}(^{13}$CO$)/$T$_{mb}($C$^{18}$O$)$ line ratio found at A$_V\gtrsim$~4 in Fig.~\ref{fig:abundances} (lower panel). Their canonical abundance ratio is recovered when each of these lines (in particular the optically thick $^{13}$CO) are corrected by their corresponding opacity (i.e. T$_{mb,corr}=\frac{\tau_0}{1-exp(-\tau_0)}\times$~T$_{mb}$; \citet{GOL99}). At the opposite end, the selective photodissotiation of the less abundant C$^{18}$O molecules is most likely responsible for the observed increase on the T$_{mb}(^{13}$CO$)/$T$_{mb}($C$^{18}$O$)$ line ratio below A$_V<$~3 mag \citep[see ][for a discussion]{VDH88}. 
As detailed in Sect.~\ref{sec:opacity}, these saturation effects also produce a non-negligible contribution to the observed linewidths of optically thick tracers like $^{13}$CO.

Although sporadically found at lower extinctions, the C$^{18}$O emission is primarily detected at A$_V\gtrsim3-4$ mag.
Supplemented by our $^{13}$CO spectra down to A$_V\gtrsim2$ mag (see Fig.~\ref{fig:kinematics}, top panel), the current sensitivity of our APEX observations limits our observation to equivalent total gas column densities of N(H)~$\gtrsim3.8\times 10^{21}$~cm$^{-2}$ \citep[assuming a standard conversion factor of N(H)(cm$^{-2})=1.9\times 10^{21}$~A$_V$ (mag);][]{SAV77}. This detection threshold corresponds to the areas within the first contour of the extinction map presented in Fig.~\ref{fig:spectra} (Left).  According to \citet{KAI14}, $\sim$~50\% of the total mass of the cloud (defined within a total extinction contour of A$_V\sim 1.5$ or N(H)$=2.9\times 10^{21}$~cm$^{-2}$) is contained at these column densities. Due to the limited coverage of our survey, the above value is assumed as an upper limit of the cloud mass fraction traced by our CO observations.

%______________________________________________ 
%______________________________________________ 

\subsection{A 6-pc-long, sonic filament}\label{sec:results}

% Velocity
Figure~\ref{fig:kinematics} (third panel) presents the velocity variations along the main axis of Musca determined from the velocity centroid (V$_{lsr}$) of the  $^{13}$CO and the C$^{18}$O (2-1) lines. 
Statistical values of $\left<|V_{lsr}(C^{18}O)-V_{lsr}(^{13}CO)|\right>=0.08\pm0.07$~km~s$^{-1}$, where 68\% of these points with absolute differences within the velocity resolution of our spectra (i.e. $|V_{lsr}(C^{18}O)-V_{lsr}(^{13}CO)|\le0.1$~km~s$^{-1}$), denote the close correlation that exist between the velocity fields traced by our CO data in those positions where the emission of both isotopologues are detected.
Despite some exceptional regions identified in the double-peaked spectra described before, the velocity structure of Musca is characterized by a continuous and unique velocity component.
At large scales, Musca presents a longitudinal and smooth south-north velocity gradient corresponding to $\nabla V_{lsr} |_{global}\sim$~0.3 km~s$^{-1}$~pc$^{-1}$. 
With similar values to those previously reported from the study of the $^{12}$CO (1--0) emission at large scales by \citet[][]{MIZ01} ($\nabla V_{lsr} |_{global}(^{12}CO)\sim$~0.2 km~s$^{-1}$~pc$^{-1}$), these global velocity gradient seems to originate in the diffuse gas, being inherited by this filament during its formation.
Locally, and at subparsec scales, the velocity field of Musca is dominated by higher magnitude velocity gradients with values of $\nabla V_{lsr} |_{local}\lesssim$~3 km~s$^{-1}$~pc$^{-1}$. Some of these local gradients appear as oscillatory motions associated with the positions of the most prominent cores in Musca \citep[e.g.; Cores 4 \& 5 in ][]{KAI14} (see also Sect.~\ref{sec:SF}). With characteristic amplitudes of $\sim$~0.25~km~s$^{-1}$, these wavy velocity excursions resemble the streaming motions associated with the formation of cores within the L1517 filaments \citep{HAC11}.

% Widths
The third quantity determined in our Gaussian fits is the full width half maximum of the line ($\Delta V$).
For each molecular species {\it i} (either $^{13}$CO or C$^{18}$O), the non-thermal velocity dispersion along the line of sight ($\sigma_{NT}$) can be directly obtained from this observable after subtracting in quadrature the thermal contribution to the line broadening as:
\begin{equation}\label{eq:NT}
	\sigma_{NT,i}=\sqrt{\frac{\Delta V_i^2}{8 ln2}-\frac{k\mathrm{T}_K}{m_i}}
\end{equation} 
where $\sigma_{th}=\sqrt{\frac{k\mathrm{T}_K}{m}}$ defines the thermal velocity dispersion of each of these tracers with a molecular weight $m$, determined by the gas kinetic temperature T$_K$.
Measurements of $\sigma_{NT}$ are commonly expressed in units of the H$_2$ sound speed (i.e., $\sigma_{th}(H_2,10~K)=c_s=$~0.19~km~s$^{-1}$, for $\mu=m(H_2)$=2.33). This ${\cal M}=\sigma_{NT}/c_s$ ratio defines the observed Mach number ${\cal M}$ classically used to distinguished between the sonic (${\cal M}\le1$), tran-sonic ($1<{\cal M}\le 2$), and supersonic (${\cal M}>2$) hydrodynamical regimes in isothermal, non-magnetic fluids.

The different values of the non-thermal velocity dispersion measured along the Musca cloud in both $^{13}$CO and C$^{18}$O lines are presented in Figure~\ref{fig:kinematics} (fourth panel) in units of the sound speed c$_s$ at T$_K=$~10~K.
As seen there, each of the CO isotopologues exhibits a roughly constant $\sigma_{NT}$ along the whole region sampled in our survey. 
A mean tran-sonic value of $\left<\sigma_{NT}/c_s\right> =1.45$ is obtained from the direct measurements of the $^{13}$CO linewidths (see also Sect.~\ref{sec:opacity}).
More extreme is the case of the C$^{18}$O (2-1) lines. With $\left<\sigma_{NT}/c_s\right>=0.73$, the emission detected in this last isotopologue presents an overwhelming fraction of $\sim$~85\% of subsonic points and a maximum transonic velocity dispersion of $\sigma_{NT}/c_s=1.7$. 

% Comparison
The quiescent properties of the gas velocity field of Musca (i.e. continuity, internal gradients, oscillatory motions, and velocity dispersions) mimic the kinematic properties found in the so-called velocity-coherent fibers, previously identified in Taurus \citep{HAC11,HAC13}. Nevertheless, and in comparison to the 0.5~pc long fibers, the velocity-coherence of Musca extents up to scales comparable to the total size of the cloud. With a total length of 6.5~pc, Musca is, therefore, the largest sonic-like, velocity-coherent structure identified in the ISM so far.

\subsection{Opacity broadening: observed linewidths and true velocity dispersions}\label{sec:opacity}

The so-called curve of growth of a molecular line describes the relationship between the intrinsic ($\Delta V_{int}$) and the observed ($\Delta V$) linewidths as a function of its central optical depth $\tau_0$ \citep[e.g.,][]{PHI79}: 
\begin{equation}\label{eq:broad}
	\beta_\tau=\frac{\Delta V}{\Delta V_{int}} =\frac{1}{\sqrt{ln 2}}\left[ln \left( \frac{\tau_0}{ln\left(\frac{2}{exp(-\tau_0)+1} \right) } \right) \right]^{1/2}
\end{equation}
This equation asymptotically converges to $\beta_\tau=\sqrt{\frac{ln\ \tau_0}{ln\ 2}}$ for $\tau_0\gg 1$.
Using Eq.~\ref{eq:broad} it is easy to prove how this opacity broadening dominates the observed linewidths for $\tau_0>10$ ($\ge$~50\%). While its effects can be neglected in the case of optically thin lines like C$^{18}$O ($\tau_0<1$ \citet{VIL94}), this opacity broadening significantly contributes to the observed linewidths of the low-J $^{13}$CO ($1\lesssim\tau_0<10$) and, in particular, $^{12}$CO transitions ($\tau_0\gtrsim100$). Thus, these opacity effects must be properly subtracted in order to compare the gas kinematics traced by the different CO isotopologues  \citep[see a detailed discussion in ][]{HAC15b}. 

Figure~\ref{fig:kinematics} displays the non-thermal velocity dispersions $\sigma_{NT}$ and $\sigma_{NT,corr}$ obtained from both observed $\Delta V$ (fourth panel) and intrinsic  $\Delta V_{int}$ linewidths (i.e. according to Eqs.~\ref{eq:NT}  and \ref{eq:broad}; fifth panel), respectively. We carried out this comparison where the line opacities of the C$^{18}$O data were available. Otherwise, the observed $\sigma_{NT}$ for $^{13}$CO were assumed as an upper limit of the true velocity dispersion of the gas. Statistically speaking, the mean contribution of the opacity broadening is estimated to be less than 5\% in the case of the optically thin C$^{18}$O linewidths and in $\sim$~25\% for the moderately opaque $^{13}$CO lines.
Remarkably, and after these opacity corrections, the gas traced in $^{13}$CO in different subregions within the Musca filament (e.g., $1.5\le L_{fil} (pc) \le 4.8$) present sonic-like velocity dispersions with $\langle\sigma_{NT,corr}$($^{13}$CO)/c$_s\rangle$=1.0.  
Overall, the opacity corrected $^{13}$CO lines still present a larger non-thermal velocity dispersion than their corresponding C$^{18}$O counterparts, indicative of an increase of the turbulent motions towards the cloud edges at A$_V<$~3-4~mag. However, and while measurable differences are still present in the gas kinematic traced by each of these CO lines, their discrepancies are restricted to changes of less than a factor of two in absolute terms and within the (tran-)sonic regime. 
  
According to Eq.~\ref{eq:broad}, these opacity broadening effects account for $\gtrsim$~60\% of the observed linewidths in the case of highly opaque lines like the low-J $^{12}$CO transitions ($\tau_0\gtrsim$~100). Values in the range of $\Delta $V($^{12}$CO(2-1))$\sim$0.9-1.3~km~s$^{-1}$ are then estimated for the expected $^{12}$CO lines in Musca with intrinsic non-thermal velocity dispersions of $\langle\sigma_{NT,corr}$/c$_s\rangle$=0.5-1.0, similar to the values deduced from our C$^{18}$O and $^{13}$CO spectra. Thus, the opacity line broadening might be also responsible for part of the observed $^{12}$CO(1-0) linewidths reported in previous molecular studies along the main axis of this cloud with $\Delta V(^{12}$CO(1-0))=1.0-1.5~km~s$^{-1}$  \citep[][H. Yamamoto, private communication]{ARN93,MIZ01}.

\subsection{Microscopic vs. macroscopic motions}\label{sec:motions}

As discussed by \citet{LAR81}, the total internal velocity dispersion of a cloud is determined by the combined contribution of the large-scale velocity variations and both the small-scale non-thermal and the thermal motions.
The simple kinematic and geometrical properties of Musca allow us to isolate and study each of these components individually.
The first two can be directly parametrized from the values measured in both the dispersion of the line velocity centroids $\sigma(V_{lsr})$ and the non-thermal velocity dispersions $\sigma_{NT}$. With a roughly constant temperature distribution according to Sect.~\ref{sec:results}, we then obtain values of $\{\sigma(V_{lsr}),\sigma_{NT}, \sigma_{th}(H_2)\}=\{1.9,0.7,1.0\}\times c_s$ from the C$^{18}$O (2-1) and $\{2.2,1.5,1.0\}\times c_s$ from the $^{13}$CO (2-1) lines (without opacity corrections). While in all the cases the observed dispersions are typically (tran-)sonic (i.e. $\lesssim 2 c_s$), the comparison of these individual components indicates that the global dispersion inside Musca cloud is dominated by the contribution of the macroscopic velocity variations along this cloud.

The nature of these macroscopic motions is also revealed by the continuity of the velocity field observed in Musca.
As seen in Fig.~\ref{fig:kinematics} (third panel), the total velocity dispersion $\sigma(V_{lsr})$ results from the combination of both local velocity excursions plus large-scale and global motions along the main axis of this filament (see also Sect.~\ref{sec:SF}). 
Compared to the random microscopic velocity variations (e.g. thermal velocity dispersion), our observations demonstrate that the largest velocity differences observed in Musca should be attributed to the presence of systematic and ordered (i.e. highly anisotropic) motions inside this cloud.

\subsection{Structure function in velocity: local oscillations vs. large-scale gradients}\label{sec:SF}

\begin{table*}[]
\caption{Model description and parameters}\label{table:model_par}
\begin{center}
\begin{tabular}{c|c|c|c|c}
	\hline
	Model & Type  & Description 	&  Velocity field 	& Parameters$^{(1)}$ \\
	\hline
	1 & Filament & Simple Oscillation  &   $V_1(L)=Asin(2\pi L / T)$ &  A = 0.25 km~s$^{-1}$, T=0.5~pc \\
	2 & Filament & Linear gradient  &   $V_2(L)=\nabla V_{lsr}\cdot L$ &  $\nabla V_{lsr}=0.25$~km~s$^{-1}$~pc$^{-1}$ \\
	3 & Filament & Linear grad. + Simple Osc. &   $V_3(L)=\nabla V_{lsr}\cdot L + Asin(2\pi L / T)$ &  Similar to models 1 \& 2 \\
	4 & Filament & Linear grad. + Complex Osc. &   $V_4(L)=\nabla V_{lsr}\cdot L + Asin(2\pi L / T(L))$ & Similar to models 1 \& 2  \\
\end{tabular}
\end{center}
\footnotesize {(1) Values refer to the representative examples used for Models 1-4 in Fig.~\ref{fig:toymodel}. 
}
\label{default}
\end{table*}

The macroscopic motions inside clouds are primarily determined by the combination of velocity oscillations and large-scale velocity gradients. In this section we aim to quantify their relative contribution to the total velocity field in Musca from the analysis of the variations of the line-centroids using the velocity structure function: $S_p(L)=\left<|v(r)-v(r+L)|^p\right>$. 
Among other correlation techniques \citep[see ][ for a review]{OSS02}, the structure function is used in molecular lines studies as a diagnostic of the velocity coherence at a given scale L \citep[e.g.,][]{MIE94}.
In practice, the structure function is commonly reframed as its {\it p}th root (i.e. $S_p(L)^{1/p}$) and described by its power-law dependency $\propto\ L^\gamma$ \citep[see][]{HEY04}.
The square root of the second-order structure function (i.e. $S_2(L)^{1/2}$) then yields
\begin{equation}\label{eq:SF}
	S_2(L)^{1/2}=\delta V=\left<|V_{lsr}(r)-V_{lsr}(r+L)|^2\right>^{1/2}=v_0\ L^{\gamma}
\end{equation}
where $v_0$ and $\gamma$  are the scaling coefficient and the power-law index, respectively. 
\citet{TAF14} used measurements of $\delta V$ to evaluate the velocity structure along different star-forming fibers in the B213-L1495 region. These authors demonstrated that the individual velocity-coherent fibers present flat and sonic-like structure functions (i.e.  $\gamma \sim 0$ and $v_0\sim c_s$) at scales of $\le$~0.5~pc. Intuitively, this behavior is expected as the result of the small velocity variations observed inside these fibers (i.e. $V_{lsr}(r) \sim V_{lsr}(r+L)$).

\begin{figure}[h]
   \centering
   \includegraphics[width=\columnwidth]{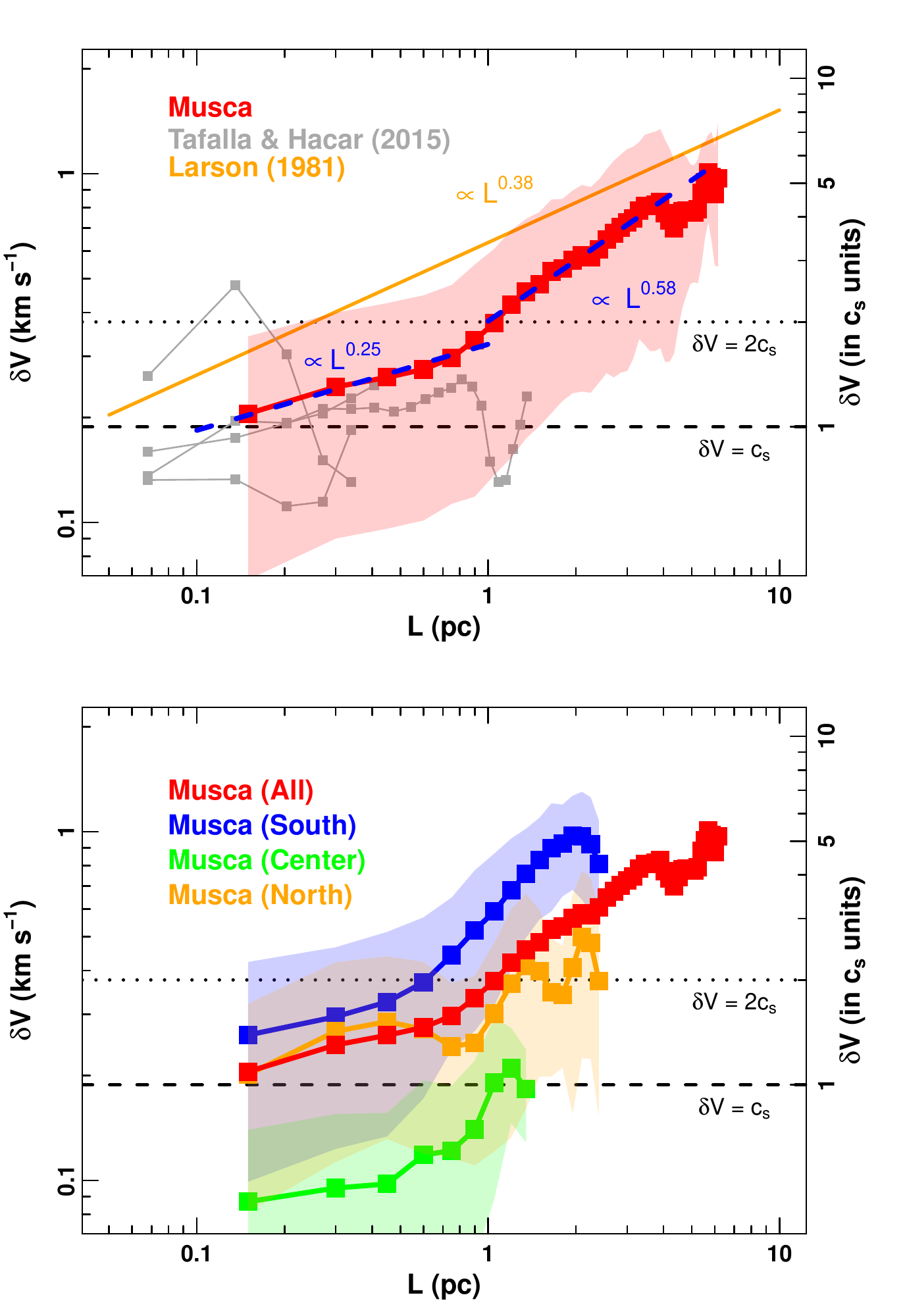}
   \caption{Second-order structure function in velocity $S_2(L)^{1/2}=\delta V$ as a function of length L (i.e. lag) derived from our $^{13}$CO observations along the main axis of Musca (red points). (Upper panel) Analysis of the global velocity field within the Musca filament. For comparison, the results obtained by \citet{TAF14} in four distinct velocity-coherent fibers in the B213-L1495 cloud are displayed in gray. The orange line denotes the Larson's velocity dispersion-size relationship with $\delta V = \sigma=0.63\ L^{0.38}$. 
 (Lower panel) Structure function of the individual south (blue), center (green), and north (orange) subregions within Musca (see text) in comparison with the total cloud (red; similar to upper panel). 
The errors on the $\delta V$ measurements are smaller than the point size. Instead, 
the shaded areas refer to the intrinsic dispersion of these measurements created by the combined contribution of the velocity excursions with different periods and the variations of the large-scale velocity gradients within this filament (see text).
}
              \label{fig:Larson}%
    \end{figure}

Compared to the short lags L sampled in the Taurus fibers, 
the study of $\delta V$ in Musca can be extended up to multiparsec scales.
Using Eq.~\ref{eq:SF}, Fig.~\ref{fig:Larson} displays the results for $\delta V$ along the main axis the Musca cloud in bins of 0.15~pc. The values presented there are calculated for all the positions detected in $^{13}$CO belonging to the main gas velocity component of this cloud and not affected by the IRAS 12322 source (see Sect.\ref{sec:fits}). Tracing the same kinematics, the use of $^{13}$CO is justified by the larger coverage and sensitivity of the emission of this tracer compared to C$^{18}$O. 
Oscillations and gradients with similar magnitudes are characteristic of the gas velocity field both parallel and perpendicular to the main axis of filaments mapped using fully sampled 2D observations \citep{HAC11}. 
A preliminary analysis of several cuts perpendicular to the axis of Musca observed in C$^{18}$O are consistent with these results \citep{HAC15}. Despite their limited coverage following the main axis of this cloud, our observations are considered a good descriptor of the gas velocity field within the gas column densities traced by our APEX observations.   

A broken power-law behavior is observed in the structure function of Musca.  
Sharing again some of the properties of the fibers, an automatic linear fit at correlation lags L~$\le$~1~pc within Musca produces a (tran-)sonic-like structure function ($\delta V \sim 1-2$~c$_s$; $v_0=0.32$~km~s$^{-1}$) with a shallow power-law dependency $\gamma = 0.25$. Conversely,  between L~>~1~pc and the completeness limit of our study at $\sim$~3~pc (or L$_{fil}/2$), the structure function becomes supersonic ($\delta V > 2$~c$_s$) showing a steeper power-law index $\gamma = 0.58$ (with $v_0=0.38$~km~s$^{-1}$). 
This rapid increase of the slope of the $\delta V$ between these two well-defined regimes
 suggests a change in the internal velocity field of the Musca cloud at characteristic scales of $\sim$~1~pc.

\begin{figure*}[ht]
   \centering
   \includegraphics[width=16cm]{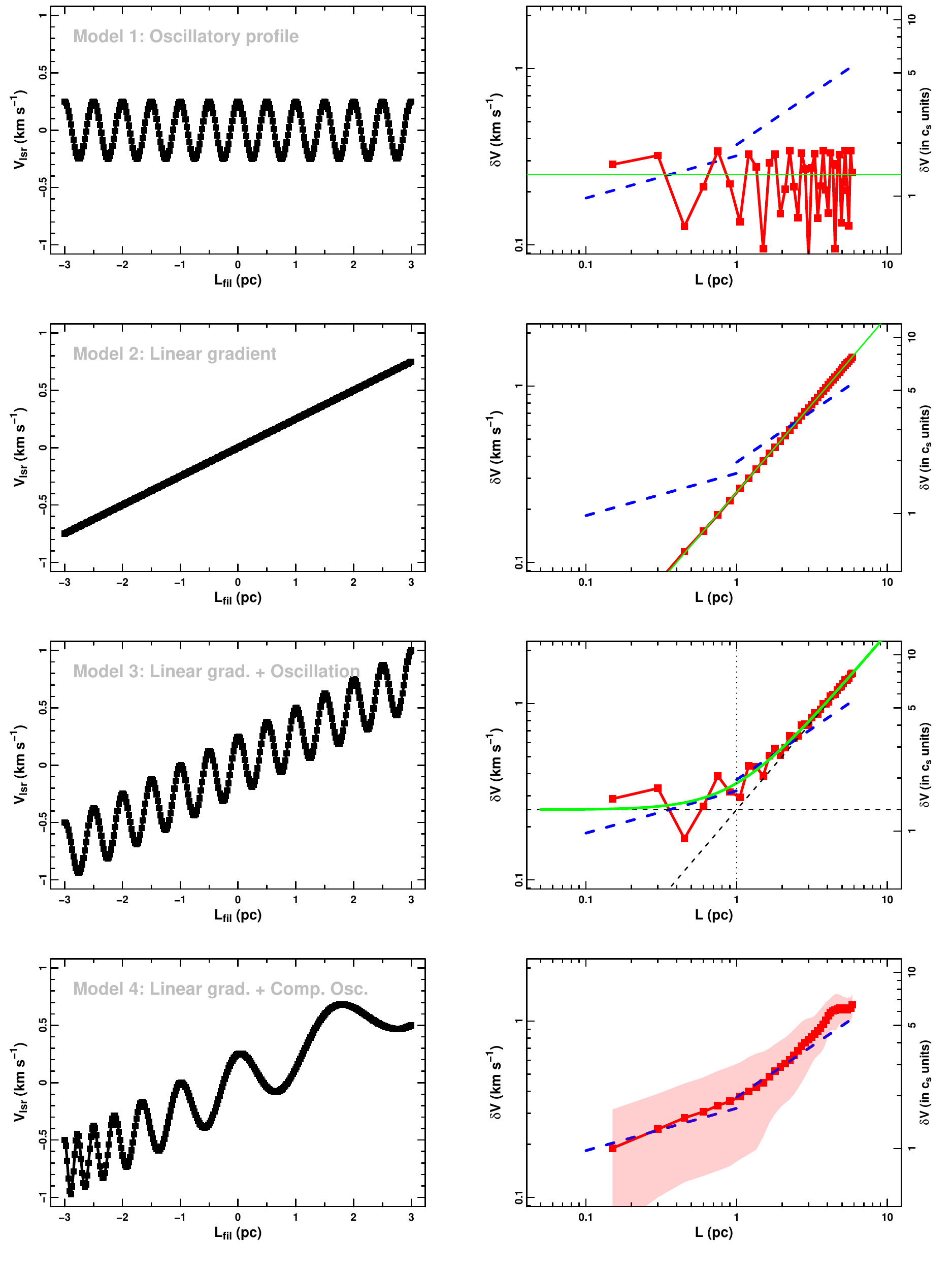}
   \caption{Position Velocity diagram (left; black dots) and its corresponding $\delta V (=S_2(L)^{1/2})$ structure function (right; red dots) for the four 1D filamentary clouds modeled in this work (see also Table~\ref{table:model_par}). From top to bottom: (Model 1) Filament presenting a sinusoidal (i.e. oscillatory) velocity profile with amplitude A=0.25~km~s$^{-1}$ and period T=0.5~pc; (Model 2) Filament presenting a longitudinal large-scale velocity gradient with $\nabla V_{lsr}=0.25$~km~s$^{-1}$~pc$^{-1}$; (Model 3) Filament presenting both a global velocity gradient and an oscillation with similar parameters to Models 1 and 2, respectively; (Model 4) Filament presenting a global velocity gradient and a complex oscillatory profile with variable period. The green lines represent the analytical fit of their corresponding structure functions (see Table~\ref{table:model_fits}).  In the case of Model 3, the vertical dashed line denotes the transitional scale where the global velocity gradient dominates over the local motions, estimated as $\Lambda \sim A/\nabla V_{lsr} $. For all plots, the blue dashed lines indicate the broken power-law dependency observed in Musca (see Fig.\ref{fig:Larson}). Similar to our observations, we note that the completeness limit of $\delta V$ is restricted to scales $\leq$~3~pc (=L$_{fil}$/2) owing to the finite size of our models. The shaded area in the last panel indicates the intrinsic dispersion per bin introduced by the presence of multiple oscillations with different periodicity in Model 4.}
              \label{fig:toymodel}
    \end{figure*}

Beyond their statistical description, the direct interpretation of spatial correlation techniques like $\delta V$ is usually hampered by the morphological complexity of the clouds and the intrinsic limitations imposed by the use of projected measurements along the line of sight \citep{SCA84}.
The geometric simplicity of Musca offers a unique opportunity to explore the origin of the power-law dependency of the structure function at different scales for the case of 1D filamentary structures. 
For that purpose, we created a set of four numerical tests to reproduce the observed $\delta V$ in filaments with different characteristic internal velocity fields (see Models 1-4 in Table~\ref{table:model_par}). In all cases, we assumed these filaments as 6~pc long, 1D structures sampled every 0.022~pc, that is, similar to our APEX observations in Musca, showing a unique and continuos velocity component. The resulting Position-Velocity diagram of each of these models are displayed in Fig.~\ref{fig:toymodel} (left). First, Models 1 and 2 describe two idealized filaments presenting a pure sinusoidal velocity field and a global velocity gradient along their main axis, respectively. These two cases describe the most fundamental velocity modes reported in the study of the internal gas kinematics inside filaments \citep{HAC11}. In addition, Model 3 is created by the linear combination of Models 1 and 2, defining a filament whose velocity field consists of a large-scale gradient superposed onto a small-scale oscillatory profile. Finally, Model 4 represents a generalized version of the previous three models where the internal kinematics of this filament is described by a linear gradient and a series of complex velocity excursions varying in period along its axis.

%______________________________________________ 
%______________________________________________ 

The resulting second-order structure function $\delta V$ derived from Eq.~\ref{eq:SF} for the four model filaments are shown in Fig.~\ref{fig:toymodel} (right). These figures reveals that each individual component of the velocity field has a particular signature in shaping their corresponding structure function. Despite their apparent complexity, in most of the cases the individual $\delta V$ can be reasonably fitted, in a first-order approximation, by simple analytic functions (see Table~\ref{table:model_fits}). 
The sinusoidal velocity excursions in Model 1 are translated into a roughly constant structure function. Different numerical tests show that the mean value of this structure function can be approximated to the amplitude of the oscillation A, that is, $\delta V_1(L)\sim A$. In the case of Model 2, its analytic solution can be easily derived from Eq.~\ref{eq:SF}, showing the linear dependency of $\delta V(L)_2$ with the global gradient $\delta V(L)_2= \nabla V_{lsr} \cdot L$. On the other hand, the relative complexity in Model 3 can be described by the addition in quadrature of the contributions of both oscillatory and linear modes as $\delta V(L)_3\sim [A^2+\nabla V_{lsr}^2 \cdot L^2]^{1/2}$. 
In these simplified simulations, the inclusion of the more complex oscillations (i.e., with different periods) in Model 4 smooth out the shape of $\delta V(L)_4$, although its average properties remain unaltered for reasonable input parameters in comparison with $\delta V(L)_3$. 
Variations on the periodicity of the oscillations and, in a more general description, changes on the velocity gradients along the main axis of these filaments introduce an intrinsic dispersion in the structure function measurements in Model 4 (shaded areas in Fig.~\ref{fig:toymodel}, lower-right panel). This intrinsic dispersion is correlated to the amplitude of the oscillations providing additional information to the velocity structure deduced from the mean values in these toy models. Due to the high accuracy of the line centroids obtained from our Gaussian fits (<0.01~km~s$^{-1}$) and the large number of pairs considered per bin (>500), the intrinsic (but still meaningful) spread of the structure function is at least an order of magnitude larger than the small nominal errors of these measurements.

\begin{table}[h]
\caption{Analytic approximation for the structure function $\delta V(L)$ of the different 1D models presented in Table~\ref{table:model_par}.}\label{table:model_fits}
\begin{center}
\begin{tabular}{c|c}
	\hline
	Model 	&   Structure Function \\
	\hline
	1$^{(1)}$ &  $\delta V(L)_1\sim A$ \\
	2 &  $\delta V(L)_2= \nabla V_{lsr} \cdot L$ \\
	3 &  $\delta V(L)_3\sim [A^2+\nabla V_{lsr}^2 \cdot L^2]^{1/2}$ \\
	4 &  --- \\
\end{tabular}
\end{center}
{\footnotesize (1) Although constant on average, the real shape of $\delta V_1$ corresponds with an oscillatory function whose nodes are produced at distances where the correlation length L coincides with a multiple of its period.}
\label{default}
\end{table}

The results obtained from the models presented in Fig.~\ref{fig:toymodel} (right) illustrate their potential use in the study of the gas dynamics inside filaments. In particular, the behavior of $\delta V(L)_3$ reveals the influence of the two fundamental velocity modes in the total velocity structure of these objects: while local velocity variations dominate the shape of $\delta V$ at short correlation lags exhibiting a flat scale dependence ($\gamma\sim 0$), the presence of global velocity gradients produces a rapid increase of its slope at larger L values ($\gamma \rightarrow 1$). Interestingly, the transition between these two regimes can be directly obtained from the parameters describing $\delta V(L)_3$, estimated at scales of $\Lambda= A/\nabla V_{lsr}$ (see also Fig.~\ref{fig:toymodel}, Model 3).   

Despite the obvious limitations of these toy models, their results closely reproduce the observed properties describing the gas velocity field within the B213-L1495 region. The internal velocity field of the fibers detected in this cloud is characterized by presenting typical velocity gradients of $\nabla V_{lsr}=0.5$~km~s$^{-1}$~pc$^{-1}$ and velocity excursions with amplitudes of A$\sim$~0.2~km~s$^{-1}$ \citep{HAC13}. Only at scales of  $\Lambda\sim A/\nabla V_{lsr}=0.2/0.5 > 0.4$~pc is the velocity field of these B213-L1495 fibers then expected to be dominated by the contribution of their global velocity gradients. These results naturally explain the roughly flat structure functions obtained for these objects up to scales of $\sim 0.5$~pc (\citet{TAF14}; see also Fig.~\ref{fig:Larson}). 
In addition, our models also predict the broken power-law behavior observed in the structure function of Musca.
As seen in Fig.~\ref{fig:toymodel}, a similar dependence is reproduced in Models 3 and 4 assuming characteristic values of $\nabla V_{lsr}=0.25$~km~s$^{-1}$~pc$^{-1}$, A=~0.25~km~s$^{-1}$ , and T=0.5~pc (see Table~\ref{table:model_par}). In this last case, the change in the slope of $\delta V(L)$ at $\Lambda= 1$~pc reveals the increasing influence of the global velocity gradient in the velocity differences observed inside this object at parsec scales. Based on these results, we then conclude that the supersonic velocity differences (or velocity dispersions) reported in the structure function of Musca are mostly created by the projection of these global motions along the main axis of this cloud.

From the analysis of its mass distribution based in extinction and continuum maps in this filament, \citet{KAI14} have suggested that Musca could be in the middle of its gravitational collapse. According to their internal substructure and number of embedded cores, these authors distinguished three subregions within this cloud: the almost unperturbed and pristine central part of Musca plus the highly fragmented north and south subregions (Fig.~\ref{fig:kinematics}, first panel). The level of fragmentation (i.e., the magnitude of the dispersion measured in the column density maps) within each of these three subregions appears to be correlated with their internal velocity dispersion and local gradients. Kainulainen et al. interpreted these properties as the development of the gravitational fragmentation progressing from the cloud edges towards the center of this filament. 

The set of parameters obtained above (A, $\nabla V_{lsr}$, and $\Lambda$) describe the average properties of the gas velocity field within the Musca filament. 
Already in our data, significant differences are also observed within the three subregions identified by \citet{KAI14} within this cloud. 
In Fig.~\ref{fig:Larson} (lower panel), we have obtained the structure function for each of these individual subregions. As for the whole cloud, we have described each of these new distributions using a broken power-law with two different slopes fitted by eye within their corresponding completeness limit.
In close agreement with the values manually derived in the Position-Velocity diagrams of Fig.~\ref{fig:kinematics}, this simple modeling captures most of the differences observed between the quiescent gas velocity field of the central region (A~=~0.08 km~s$^{-1}$, $\nabla V_{lsr}=$~0.15 km~s$^{-1}$~pc$^{-1}$, $\Lambda =$~0.5 pc), and the combination of oscillations and gradients of different magnitudes present in the adjacent south (A~=~0.25 km~s$^{-1}$, $\nabla V_{lsr}=$~0.50 km~s$^{-1}$~pc$^{-1}$, $\Lambda =$~0.5 pc) and north regions (A~=~0.22 km~s$^{-1}$, $\nabla V_{lsr}=$~0.20 km~s$^{-1}$~pc$^{-1}$, $\Lambda =$~1.1 pc). The high sensitivity of our models to these local variations denote their potential as diagnostic tools of the gas velocity structure in 1D filamentary structures. 

%______________________________________________ 
%______________________________________________ 

\section{Discussion}

\subsection{Departure from Larson's velocity dispersion-size relationship}

The so-called Larson's velocity dispersion-size relationship (often referred to as linewidth-size relationship) defines the observational correlation found between the velocity dispersion ($\sigma$) of the gas as a function of cloud size (L). Collecting data from the literature for different nearby clouds, \citet{LAR81} first pointed out that this relationship can be described by a power-law dependence  $\sigma (\mathrm{km s}^{-1})\simeq C\ L(\mathrm{pc})^\Gamma$, with coefficients C=0.63 and $\Gamma$=0.38, respectively\footnote{The original coefficient C(Larson)=1.1 given by \citet{LAR81} was derived for the total 3D velocity dispersion ($\sigma_{3D}$). To allow a direct comparison with the 1D structure function $\delta V$, the value used here corresponds to the 1D velocity dispersion, or C=$\mathrm{C(Larson)}/\sqrt{3}=$~0.63, assuming an isotropic velocity dispersion \citep[i.e., $\sigma_{1D}=\sigma_{3D}/\sqrt{3}$; ][]{MYE83}.}.
In later studies, roughly similar power-law behaviors were consistently reported for the velocity structure in both galactic \citep[C=1.0 and $\Gamma=0.5$; ][]{SOL87} and extragalactic clouds \citep[C=0.44 and $\Gamma=0.6$; ][]{BOL08} along more than three orders of magnitudes in scale between $\sim$~0.1--100~pc . The small variations found in both C and $\gamma$ parameters describing the Larson's velocity dispersion-size relationship inside molecular clouds are interpreted as an observational signature of the universality of the ISM turbulence \citep{HEY04}. 

As demonstrated by \citet{HEY04}, the universal behavior of the velocity dispersion-size relationship based on cloud-to-cloud comparisons can only be explained if each of these individual objects presents a quasi-identical and Larson-like internal structure function (i.e., $\delta V(L)_i \sim \sigma (L) \propto L^\gamma$). 
Contrary to their conclusions, and as illustrated in Fig.~\ref{fig:Larson} (Upper panel) (see also Sect.~\ref{sec:SF}), the observed power-law dependency of the $\delta V(L)$ function in Musca systematically deviates from the Larson's velocity dispersion-size predictions in both its slope and its absolute scaling coefficient, even when its intrinsic dispersion is considered. The two functions seem to converge only at large spatial correlation lags L>3~pc. In Musca, however, these large velocity differences are produced by the presence of large-scale and ordered velocity gradients instead of by turbulent random motions (Sect.~\ref{sec:SF}). 

The flat structure function reported for the internal gas kinematics inside fibers at scales of $\sim$~0.5~pc was used by \citet{TAF14} to prove the particular nature of these objects fully decoupled from the turbulent cascade. A similar conclusion can be drawn from the analysis of the $\delta V(L)$ dependence observed in Musca at scales comparable with its total $\sim$~6.5~pc length. 
Large-scale departures from Larson's predictions were previously suggested for individual filamentary objects like the Ophiuchus Streamers \citep{LOR89}. 
The detection of quiescent clouds like Musca indicate that some of these filaments could present an internal velocity dispersion deviating from Larson's velocity dispersion-size relationship at parsec-scales. 
Without necessarily contradicting the global gas properties described by this last empirical relationship, our observations show that the local kinematics of some individual subregions within molecular clouds might not necessarily follow Larson's predictions.

\subsection{Origin of the Musca cloud}

The results presented in this paper extend the size of the sonic regime of the gas inside filaments by about an order of magnitude compared to previous studies. Low-amplitude continuous longitudinal gradients seemed to be characteristic of fibers. However, the presence of similar velocity-coherent, sonic-like structures was thought to be restricted to subparsec scales \citep{HAC11,HAC13}. The low velocity dispersion (Sect.\ref{sec:results}), as well as the shallow dependency of its structure function (Sect.\ref{sec:SF}) indicate that no significant internal supersonic, turbulent motions are present in Musca at multiparsec scales.

The dynamical properties of Musca make its origin puzzling.
Pure hydrodynamical simulations typically produce dense filamentary structures with low velocity dispersions in the post-shock regions after the collision of supersonic flows \citep[e.g.,][]{PAD01}. The almost ideal conditions required to produce an isolated, unperturbed parsec-scale filamentary structure like Musca in this shock-dominated scenario could explain the uniqueness of this object.
Its elongated nature could also be favored by the presence of the highly ordered magnetic field structure reported in previous studies \citep{PER04}. The orthogonal orientation of the magnetic field lines with respect to the main axis of Musca could promote the gravitational condensation of material towards its axis. According to MHD simulations \citep[e.g.,][]{NAK08}, this configuration would also inhibit the gas motions across the magnetic field lines, preventing or delaying the internal fragmentation along this cloud and leading to a slow and inefficient star formation inside Musca, in agreement with our observational results. These different formation scenarios will be explored with the analysis of the gas velocity field perpendicular to this filament in a future paper \citep{HAC15}.

Our findings in Musca could potentially open a new window in the study of the evolution of filaments in molecular clouds.
The intricate nature of more massive star-forming filaments, like B213-L1495 \citep{SCH10,HAC13}, complicates the detailed analysis of these objects and, in particular, their comparison to simulations. 
Far from this complexity, Musca could be used as benchmark of more simplified physical models (e.g., see Sect.\ref{sec:SF}). 
Thanks to its favorable structure, geometry, and kinematic properties, direct predictions for distinct key processes like the fragmentation and stability mechanisms in filaments could easily be tested in this cloud \citep[e.g.,][]{KAI14}. 
All these extraordinary properties make Musca an ideal laboratory for studying the nature and evolution of filamentary clouds.

\section{Conclusions}

We investigated the dynamical state of the Musca cloud using APEX submillimeter observations. We sampled 300 positions along the main axis of this filament in both $^{13}$CO and C$^{18}$O (2-1) lines. We characterized the internal kinematic structure of Musca from the analysis of the different line profiles. The main conclusions of this work are as follows:
\begin{enumerate}
	\item The velocity structure of the Musca cloud is characterized by a continuous and quiescent velocity field along its total length of 6~pc. Its internal gas kinematics traced by the C$^{18}$O (2-1) emission is mainly described by a single velocity component presenting (tran-)sonic non-thermal velocity dispersions (i.e., $\left<\sigma_{NT}/c_s\right> \lesssim 1$) and smooth and oscillatory-like velocity profile with internal velocity gradients of $\nabla V_{lsr}\lesssim$~3 km~s$^{-1}$~pc$^{-1}$). Consistent results are obtained from the analysis of the $^{13}$CO (2-1) lines.
	\item The properties of the gas velocity field in Musca present similar characteristics to those reported in previous studies for the so-called velocity-coherent fibers at scales of $\sim$~0.5~pc. About an order of magnitude larger in extension, Musca is therefore the largest sonic-like structure identified so far in nearby clouds.
	\item The analysis of local and global motions indicates that the largest velocity variations inside Musca arise from the parsec-scale projection of a global velocity gradient along its main axis. Hence, the velocity field in this cloud is dominated by macroscopic, ordered, and systematic motions.
	\item We have quantified the macroscopic internal motions within the Musca cloud using the second-order structure function ($\delta V$). 
	Applied to the line velocity centroids of our CO observations, this function presents a broken power-law behavior ($\delta V\propto$~L$^{\gamma}$) showing a sharp change in its slope from $\gamma=0.25$ to $\gamma=0.58$ at scales of $\sim$~1~pc. A simple 1D modeling demonstrates that both the different slopes and the position of this reported knee in the $\delta V$ function are consistent with the expected structure function of a filament with an internal velocity field described by a global velocity gradient and a series of local velocity oscillations, in agreement with the observations.
	\item Within the completeness limit of our study ($\lesssim$~3~pc), the parameters describing $\delta V$ reported for Musca are systematically lower than the velocity dispersions expected from the canonical Larson's velocity dispersion-size relationship at similar scales. This departure from Larson's predictions suggests the existence of parsec-scale, sonic-like structures fully decoupled from the supersonic turbulent regime in the ISM.
	\end{enumerate}

\begin{acknowledgements}
      We thank Palle Moller, Carlos DeBreuck, and Thomas Stanke for their help developing the new observing technique used in these APEX observations.
      The authors also thank Akira Mizuno, Hiroaki Yamamoto, and Yasuo Fukui for kindly providing their Musca NANTEN $^{12}$CO (1--0) data.
      A.H. gratefully acknowledges support from Ewine van Dishoeck, John Tobin, Magnus Persson, and Sylvia Ploeckinger during his stay at the Leiden Observatory. 
      A.H. thanks the insightful discussions and comments from Andreas Burkert, Jan Forbrich, Paula Teixeira, and Oliver Czoske.
      J.K. was supported by the Deutsche Forschungsgemeinschaft priority program 1573 (``Physics of the Interstellar Medium''). 
      J.K. gratefully acknowledges support from the Finnish Academy of Science and Letters/V\"ais\"al\"a Foundation. 
      This publication is supported by the Austrian Science Fund (FWF).
      This research has made use of the SIMBAD database, operated at CDS, Strasbourg, France.
      This research made use of the TOPCAT software \citep{TAY05}.
\end{acknowledgements}

%-------------------------------------------------------------------


\begin{thebibliography}{}

\bibitem[Andr{\'e} et al.(2010)]{AND10} Andr{\'e}, P., Men'shchikov, A., Bontemps, S., et al.\ 2010, \aap, 518, LL102 

\bibitem[Andr{\'e} et al.(2014)]{AND14} Andr{\'e}, P., Di Francesco, J., Ward-Thompson, D., et al.\ 2014, Protostars and Planets VI, 27 

\bibitem[Arnal et al.(1993)]{ARN93} Arnal, E.~M., Morras, R., \& Rizzo, J.~R.\ 1993, \mnras, 265, 1

\bibitem[Arzoumanian et al.(2013)]{ARZ13} Arzoumanian, D., Andr{\'e}, P., Peretto, N., Konyves, V.\ 2013, \aap, 553, AA119 

\bibitem[Benson \& Myers(1989)]{BEN89} Benson, P.~J., \& Myers, P.~C.\ 1989, \apjs, 71, 89 

\bibitem[Bolatto et al.(2008)]{BOL08} Bolatto, A.~D., Leroy, A.~K., Rosolowsky, E., Walter, F., \& Blitz, L.\ 2008, \apj, 686, 948 

\bibitem[Cazzoli et al.(2003)]{CAZ03} Cazzoli, G., Puzzarini, C., \& Lapinov, A.~V.\ 2003, \apjl, 592, L95 

\bibitem[Cazzoli et al.(2004)]{CAZ04} Cazzoli, G., Puzzarini, C., \& Lapinov, A.~V.\ 2004, \apj, 611, 615 

\bibitem[Elmegreen \& Scalo(2004)]{ELM04} Elmegreen, B.~G., \& Scalo, J.\ 2004, \araa, 42, 211 

\bibitem[Evans(1999)]{EVA99} Evans, N.~J., II 1999, \araa, 37, 311 

\bibitem[Franco(1991)]{FRA91} Franco, G.~A.~P.\ 1991, \aap, 251, 581 

\bibitem[Feitzinger \& Stuewe(1984)]{FEI84} Feitzinger, J.~V., \& Stuewe, J.~A.\ 1984, \aaps, 58, 365 

\bibitem[Frerking et al.(1982)]{FRE82} Frerking, M.~A., Langer, W.~D., \& Wilson, R.~W.\ 1982, \apj, 262, 590 

\bibitem[Goldsmith \& Langer(1999)]{GOL99} Goldsmith, P.~F., \& Langer, W.~D.\ 1999, \apj, 517, 209 

\bibitem[Goodman et al.(1998)]{GOO98} Goodman, A.~A., Barranco, J.~A., Wilner, D.~J., \& Heyer, M.~H.\ 1998, \apj, 504, 223 

\bibitem[Hacar \& Tafalla(2011)]{HAC11} Hacar, A., \& Tafalla, M.\ 2011, \aap, 533, A34 

\bibitem[Hacar et al.(2013)]{HAC13} Hacar, A., Tafalla, M., Kauffmann, J., \& Kov{\'a}cs, A.\ 2013, \aap, 554, A55 

\bibitem[Hacar et al.(2015a, submitted)]{HAC15b} Hacar, A., Alves, J., Burkert, A., \& Goldsmith, P., \ 2015, submitted to A\&A.

\bibitem[Hacar et al.(2015b, in prep)]{HAC15} Hacar, A., Kainulainen, J., Beuther, H., Tafalla, M., \& Alves, J.\ 2015, in prep.

\bibitem[Heyer \& Brunt(2004)]{HEY04} Heyer, M.~H., \& Brunt, C.~M.\ 2004, \apjl, 615, L45 

\bibitem[Juvela et al.(2010)]{JUV10} Juvela, M., Ristorcelli, I., Montier, L.~A., et al.\ 2010, \aap, 518, L93 

\bibitem[Juvela et al.(2012)]{JUV12} Juvela, M., Ristorcelli, I., Pagani, L., et al.\ 2012, \aap, 541, AA12 

\bibitem[Kainulainen et al.(2009)]{KAI09} Kainulainen, J., Beuther, H., Henning, T., \& Plume, R.\ 2009, \aap, 508, L35 

\bibitem[Kainulainen et al.(2014)]{KAI14b} Kainulainen, J., Federrath, C., \& Henning, T.\ 2014, Science, 344, 183 

\bibitem[Kainulainen et al.(2015)]{KAI14}  Kainulainen, J., Hacar, A., Alves, J., et al.\ 2015, arXiv:1507.03742 

\bibitem[Larson(1981)]{LAR81} Larson, R.~B.\ 1981, \mnras, 194, 809

\bibitem[Loren(1989)]{LOR89} Loren, R.~B.\ 1989, \apj, 338, 925 

\bibitem[McKee \& Ostriker(2007)]{MAC07} McKee, C.~F., \& Ostriker, E.~C.\ 2007, \araa, 45, 565 

\bibitem[Miesch \& Bally(1994)]{MIE94} Miesch, M.~S., \& Bally, J.\ 1994, \apj, 429, 645 

\bibitem[Mizuno et al.(2001)]{MIZ01} Mizuno, A., Yamaguchi, R., Tachihara, K., et al.\ 2001, \pasj, 53, 1071 

\bibitem[Myers et al.(1983a)]{MYE83a} Myers, P.~C., Linke, R.~A., \& Benson, P.~J.\ 1983, \apj, 264, 517

\bibitem[Myers(1983b)]{MYE83} Myers, P.~C.\ 1983, \apj, 270, 105

\bibitem[Nakamura \& Li(2008)]{NAK08} Nakamura, F., \& Li, Z.-Y.\ 2008, \apj, 687, 354 

\bibitem[Ossenkopf \& Mac Low(2002)]{OSS02} Ossenkopf, V., \& Mac Low, M.-M.\ 2002, \aap, 390, 307 

\bibitem[Padoan et al.(2001)]{PAD01} Padoan, P., Juvela, M., Goodman, A.~A., \& Nordlund, {\AA}.\ 2001, \apj, 553, 227 

\bibitem[Pereyra \& Magalh{\~a}es(2004)]{PER04} Pereyra, A., \& Magalh{\~a}es, A.~M.\ 2004, \apj, 603, 584 

\bibitem[Phillips et al.(1979)]{PHI79} Phillips, T.~G., Huggins, P.~J., Wannier, P.~G., \& Scoville, N.~Z.\ 1979, \apj, 231, 720 

\bibitem[Pineda et al.(2010)]{PIN10} Pineda, J.~E., Goodman, A.~A., Arce, H.~G., et al.\ 2010, \apjl, 712, L116 

\bibitem[Planck Collaboration et al.(2014)]{PLA14} Planck Collaboration, Ade, P.~A.~R., Aghanim, N., et al.\ 2014, arXiv:1411.2271 

\bibitem[Rohlfs \& Wilson(2004)]{TOOLS} Rohlfs, K., \& Wilson, T.~L.\ 2004, Tools of radio astronomy, 4th rev.~and enl.~ed., by K.~Rohlfs and T.L.~Wilson.~ Berlin: Springer, 2004,

\bibitem[Savage et al.(1977)]{SAV77} Savage, B.~D., Bohlin, R.~C., Drake, J.~F., \& Budich, W.\ 1977, \apj, 216, 291

\bibitem[Scalo(1984)]{SCA84} Scalo, J.~M.\ 1984, \apj, 277, 556

\bibitem[Schmalzl et al.(2010)]{SCH10} Schmalzl, M., Kainulainen, J., Quanz, S.~P., et al.\ 2010, \apj, 725, 1327 

\bibitem[Solomon et al.(1987)]{SOL87} Solomon, P.~M., Rivolo, A.~R., Barrett, J., \& Yahil, A.\ 1987, \apj, 319, 730 

\bibitem[Tafalla \& Hacar(2015)]{TAF14} Tafalla, M., \& Hacar, A.\ 2015, \aap, 574, AA104 

\bibitem[Taylor(2005)]{TAY05} Taylor, M.~B.\ 2005, Astronomical Data Analysis Software and Systems XIV, 347, 29  

\bibitem[van Dishoeck \& Black(1988)]{VDH88} van Dishoeck, E.~F., \& Black, J.~H.\ 1988, \apj, 334, 771

\bibitem[Vilas-Boas et al.(1994)]{VIL94} Vilas-Boas, J.~W.~S., Myers, P.~C., \& Fuller, G.~A.\ 1994, \apj, 433, 96 

\bibitem[Wilson \& Rood(1994)]{WIL94} Wilson, T.~L., \& Rood, R.\ 1994, \araa, 32, 191 

\bibitem[Zuckerman \& Palmer(1974)]{ZUC74} Zuckerman, B., \& Palmer, P.\ 1974, \araa, 12, 279 


\end{thebibliography}
\end{document}